\documentclass[useAMS,usenatbib]{mn2e}
\usepackage{amsmath}
\usepackage{graphicx}
\def\sun{\hbox{$\odot$}}
\def\mta{M_{\textrm{TA}}}
\def\mtade{M_{\textrm{TA}\Lambda}}
\def\mvir{M_{\textrm{vir}}}

\title[Weighing the Local Group in the Presence of Dark Energy]
  {Weighing the Local Group in the Presence of Dark Energy}
  \author[C. Partridge et al.]
  {C.~Partridge$^1$, O.~Lahav$^1$ and Y.~Hoffman$^2$ \\
 $^1$University College London, Gower Street, London WC1E 6BT, UK\\
 $^2$Racah Institute of Physics, Hebrew University, Jerusalem 91904, Israel }
\date{Released 2013}

\pagerange{\pageref{firstpage}--\pageref{lastpage}} \pubyear{2013}

\def\LaTeX{L\kern-.36em\raise.3ex\hbox{a}\kern-.15em
    T\kern-.1667em\lower.7ex\hbox{E}\kern-.125emX}

\begin{document}

\label{firstpage}

\maketitle

\begin{abstract}
We revise the mass estimate of the Local Group (LG) when Dark Energy (in the form of the Cosmological Constant) is incorporated into the Timing Argument (TA) mass estimator for the Local Group (LG). Assuming the age of the Universe and the Cosmological Constant according to the recent values from the Planck CMB experiment, we find the mass of the LG to be $\mtade = (4.73 \pm 1.03) \times 10^{12} M_{\sun}$ which is 13\% higher than the classical TA mass estimate. This partly explains the discrepancy between earlier results from LCDM simulations and the classical TA.  When a similar analysis is performed on 16 LG-like galaxy pairs from the CLUES simulations, we find that the scatter in the ratio of the virial to the TA estimated mass is given by $\mvir / \mtade = 1.04 \pm 0.16$. Applying it to the LG mass estimation we find a calibrated $\mvir = (4.92  \pm 1.08 \textrm{(obs.)} \pm 0.79 \textrm{(sys.)} ) \times 10^{12} M_{\sun}$.

\end{abstract}


\section{Introduction}

In this paper, we explore the possibility that Dark Energy would have an influence on  estimation of galaxy  masses via the Timing Argument (TA).   The TA relates the mass of the Local Group (LG) with the age of the universe by a method that was formulated by \cite{kahnwoltjer1959} and refined by \cite{lyndenbell1981}. Since the bulk of the mass of the LG is concentrated in M31 (Andromeda Galaxy) and the Milky Way (MW), it can be modelled as an isolated system of two point masses.  Because M31 is approaching the MW on a low angular momentum orbit, the LG can be well approximated by an isolated two-body problem with zero angular momentum (\cite{vdm12} and \cite{cosmoLG}). The Timing Argument of \cite{kahnwoltjer1959} is perhaps not sufficiently credited as one of the first indicators of dark matter (\cite{freeman2013}).

Soon after the time of the Big Bang, these two galaxies must have been in the same place, with zero separation.  The two galaxies are observed at present to be approaching each other, implying that they would have reached maximum separation at some point in the past. Because they have low orbital angular momentum, the two galaxies would be on a nearly head-on collision course. In that case, their disks would have  already been severely disrupted unless they are not currently on their first close approach. Since both galaxies have unperturbed disks, it is assumed that they are on their first passage. Therefore the mass of the MW/M31 galaxy pair can be estimated by the TA, as shown by \cite{lyndenbell1981} and \cite{binneytremaine}.

It seems that there is some confusion in the literature about the role of $\Lambda$ in the growth of structure.  When global fluctuations relative to the background are being discussed, as in \cite{peebles1980}, the $\Lambda$ term cancels out and seemingly has no effect on dynamics.  This comes from the fact that these problems are being considered in co-moving coordinates, and therefore the expansion of the universe due to Dark Energy is already captured by the background expansion of the coordinate system.

However, when local dynamics are being considered, particularly for gravitationally bound systems like the Local Group, the equations used to model said systems are usually written in physical coordinates, hence the $\Lambda$ term should be there, as in the case of  spherical collapse.  Investigation of the effects of incorporating $\Lambda$ into the TA is a good test of the assertion that $\Lambda$ should be included in models for bound systems in physical coordinates. 

The acceleration of the galaxies towards each other is described by the radial differential equation, where $r$ is a proper radial coordinate, 
\begin{equation}
\label{ODE}
\frac{d^2 r}{dt^2} = -\frac{GM}{r^2} + \frac{\Lambda}{3}r.
\end{equation}
In this equation,  $r$ represents the scalar distance, or radial separation, between the MW and M31, and $M$ is the sum of the masses for the MW and M31, although it is recognised that halo mass is an ill-defined quantity.  For the purposes of our analysis, our treatment only applies to the simplest DE model, namely one with a cosmological constant where the equation of state $w=-1$ at all times. Further, we assume that the masses of the MW and M31 are fixed in time. Tidal interactions and orbital angular momentum are also ignored.  

Equation \ref{ODE} is the same as the equation for a Lema\^itre-Tolman-Bondi spherical collapse which was applied in the past for collapse in the presence of Dark Energy (e.g. \cite{lahav1991}, \cite{wangsteinhardt}, \cite{maorlahav}). Equation \ref{ODE} was applied to the Local Group by \cite{axenides},  \cite{PP06}, \cite{PP08},  \cite{binneytremaine}, \cite{chernin2009} and \cite{chernin2006}, and to the Virgo-centric infall by \cite{hoffman2007}. We point out that Lambda was not taken into account in Local Group Timing Argument recent analyses e.g. \cite{timingargwhite} and \cite{vdm12}. The goal of this paper is investigate if the inclusion of Lambda in the TA better models the dynamics of LG-like galaxy pairs in LCDM simulations. In particular, \cite{timingargwhite} compared their LG-like galaxy pairs from their LCDM simulations with the classical TA with $\Lambda=0$, i.e.
\begin{equation}
\label{newtonianODE}
\frac{d^2 r}{dt^2} = -\frac{GM}{r^2}.
\end{equation}

This differential equation is parameterised as follows (e.g. \cite{binneytremaine} and \cite{kahnwoltjer1959}):
\begin{equation}
\label{TAeq1}
r = a(1 - \cos{\theta})
\end{equation}
\begin{equation}
\label{TAeq2}
t = \sqrt{\frac{a^3}{GM}}\,(\theta - \sin{\theta})
\end{equation}
\begin{equation}
\label{TAeq3}
v = \frac{dr}{dt} = \sqrt{\frac{GM}{a}} \frac{\sin{\theta}}{1 - \cos{\theta}}.
\end{equation}
An angle of $\theta = 0$ corresponds to the closest approach at $t=0$ (the Big Bang). The angle $\theta = \pi$ is where the maximum radial separation $2a$ occurs. For a given known radial separation $r$, velocity $v$, and the age of the universe $t$, these give the mass of the system. 

In this paper, we explore if the addition of $\Lambda$ to the LG TA could improve the agreement with the N-Body results. In order to include the effect of $\Lambda$, we sought to numerically solve Eq. \ref{ODE}, which was numerically solved with $\Lambda \neq 0$ and $\Lambda = 0$ so that we could directly compare the behaviour of the DE TA model with that of the analytic TA model.   This comparison is interesting because some discrepancy was noted by \cite{timingargwhite} between masses derived from their simulation and classical TA. Furthermore, the connection between simulations and the Newtonian Eq. \ref{ODE} is non-trivial. In the simulations, $\Lambda$  only appears in the scale factor to describe the expansion of the simulated box, while in Eq. \ref{ODE}, the Lambda term as repulsive force in proper coordinates.

The outline of the paper is as follows: In Section 2 we solve Eq. \ref{ODE} numerically so that we could directly compare the behaviour of the TA with and without Lambda, and apply both to the LG, indicating a derived mass 13\% higher from the TA with $\Lambda$ as compared to the classical TA. In Section 3, we compare the TA with $\Lambda$ against simulations with the motivation of calibrating the TA model is even after $\Lambda$ was included. Section 4 discusses the implications of our results.

\section{Application to Local Group}
 
Four input parameters are required to solve Eq. 1 for the mass $M$.  For ease of comparison with \cite{timingargwhite}, we assume at the present epoch the physical separation between MW and M31 is $r= 784 \pm 21$\,kpc (\cite{stanek1998}), and the radial velocity is $v=-130 \pm 8\,\textrm{km}\,\textrm{s}^{-1}$ (\cite{M31transverse} as used by \cite{timingargwhite}). We note that unlike our radial model,  \cite{M31transverse} assumed a model with transverse motion of M31, however we use their $v$ for simplicity.  For the cosmological parameters, we use the recent Planck CMB experiment values: the Cosmological  Constant term $\Omega_{\Lambda} = \Lambda/3H_0^2$ where $\Omega_{\Lambda} = 0.69 \pm 0.02$ and $H_0 = 67.4 \pm 1.4$\,km/sec/Mpc, and the age of the Universe is $13.81 \pm 0.06$\,Gyr (\cite{planck}).

\begin{figure} 
\centering 
\includegraphics[scale=0.5]{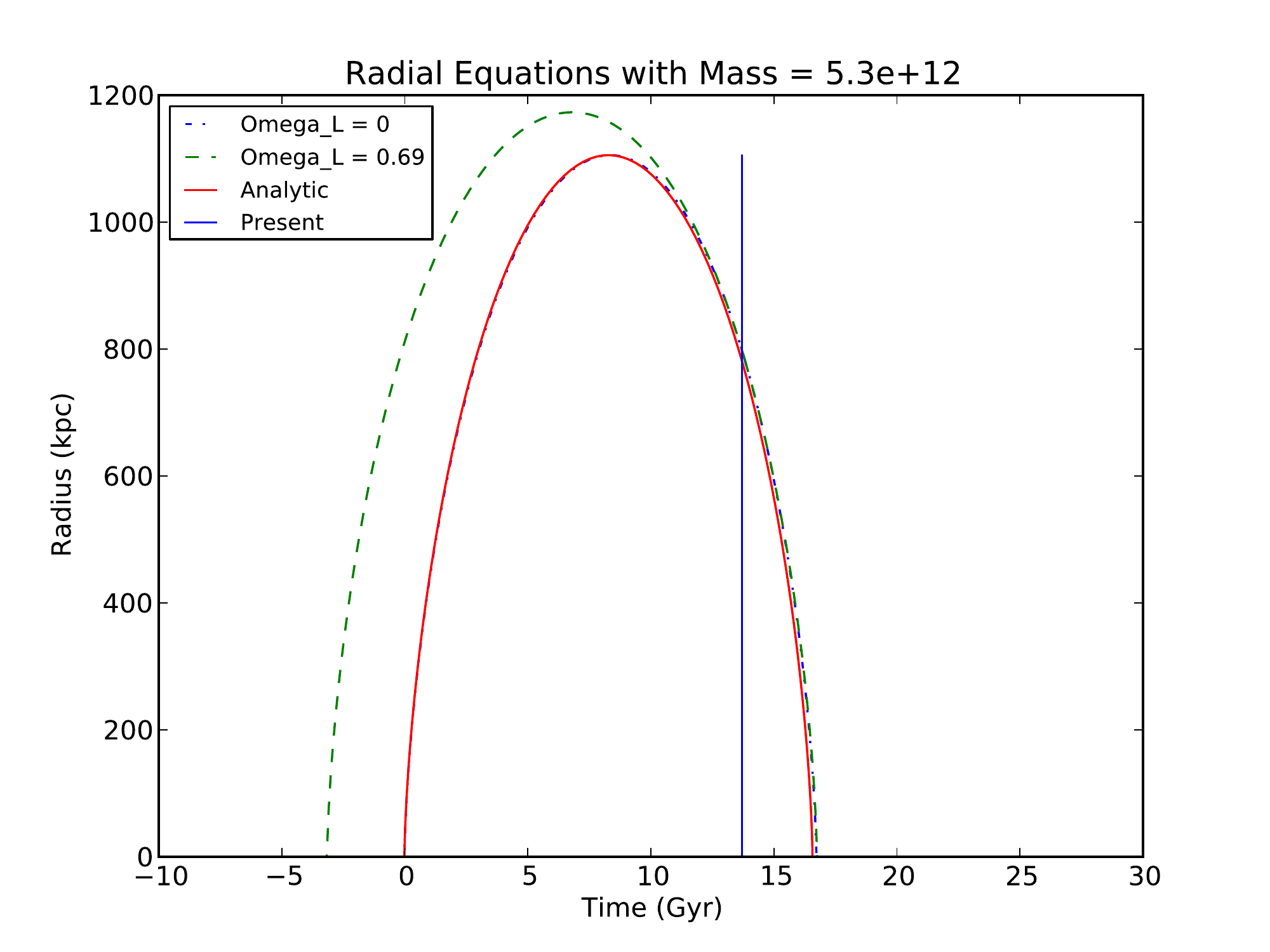}
\caption{The projected distance between MW and M31 according to Eq. \ref{ODE} for the radial motion and analytic equation for $\mta = 5.30 \times 10^{12}\,\text{M}_{\sun}$. We see that the traditional analytic solution agrees with the numerical solution for the case where $\Lambda = 0$. In the case where $\Omega_{\Lambda} = 0.69$, we find a longer age of the Universe (see text for explanation).}
\label{fig:radialobs}  
\end{figure}
\begin{figure} 
\centering 
\includegraphics[scale=0.5]{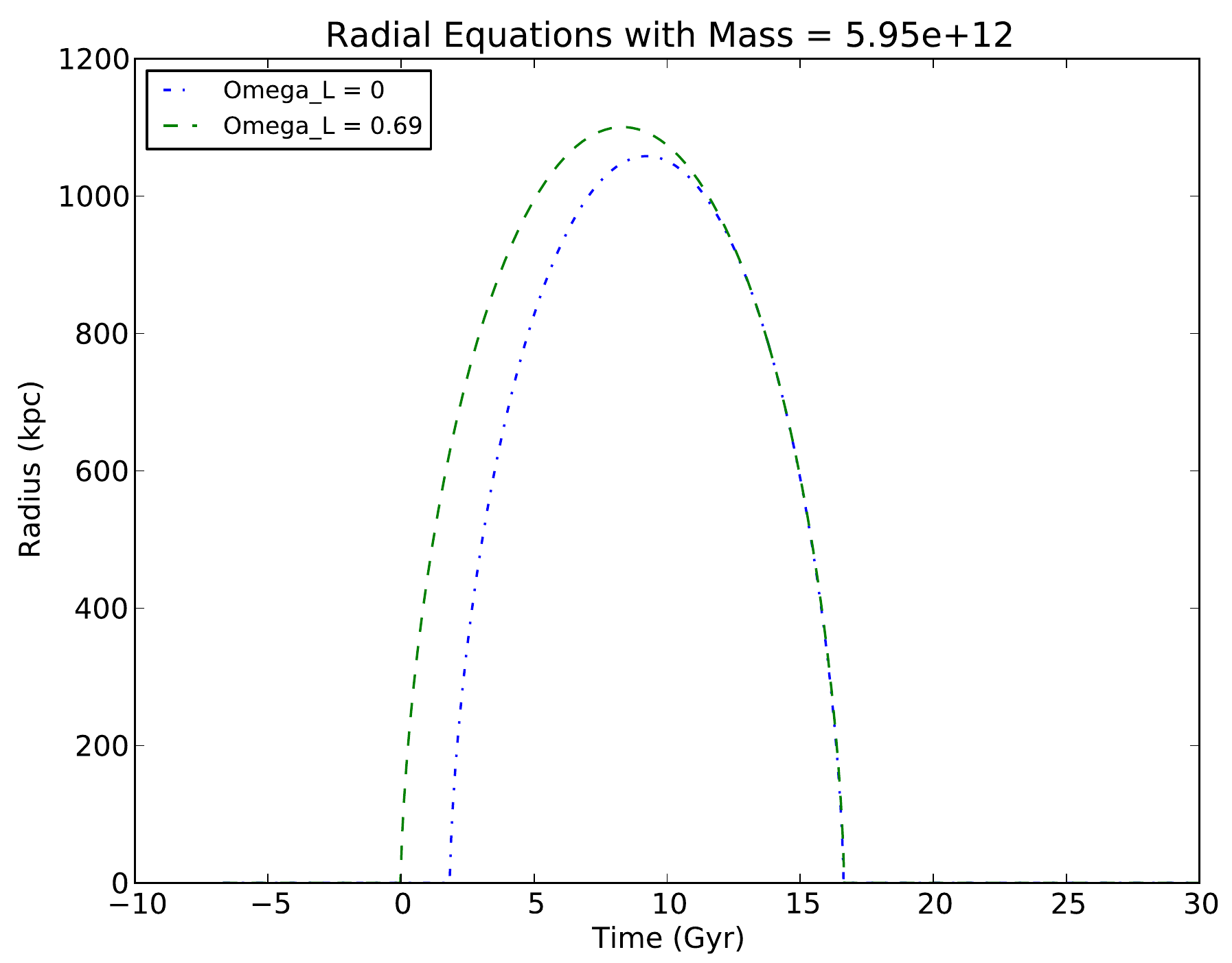}
\caption{The radial equation including the Dark Energy term has $r=0$ at $t=0$ at $\mtade = 5.95 \times 10^{12}\,\text{M}_{\sun}$, which is the value that results from the observed $r, v, t$, and $\Lambda$, as given in the text. In the case where $\Omega_{\Lambda} = 0.69$, we now find that the age of the Universe has shortened compared with Fig. \ref{fig:radialobs}.}
\label{fig:radialshift}  
\end{figure}

It can be seen from Fig. \ref{fig:radialobs} that both the analytic and non-Lambda $r(t)$ curves are identical and that they converge to the expected radius of $r = 0$ when time $t = 0$, i.e. that the MW and M31 were coincident at the beginning of the universe, as the TA assumes.  For the classical TA, we derive a Timing Argument mass of $\mta = (5.30 \pm 0.47) \times 10^{12}\,\text{M}_{\sun}$, where the statistical error was estimated by propagating the above errors on input parameters through the TA mass model and then adding them in quadrature. However, in the presence of Dark Energy for the same mass, the curve $r(t)$ reaches zero 4 Gyr before the Big Bang.  According to the TA, there are two points on the curve where the radial separation and velocity are known: at the beginning of the universe, and at the present time.  Therefore,  we have to find the mass which forces the curve $r(t)$ in the presence of Dark Energy to begin at $r=0$ at $t=0$ (see Fig.\,\ref{fig:radialshift}).  By an iterative process, we derived $\mtade = (5.95 \pm 0.52) \times 10^{12}\,\mathrm{M_{\sun}}$. This shows that the Dark Energy term would have a significant effect on TA mass estimates for the LG, increasing it by 12\%. 

\begin{figure}  
\centering 
\includegraphics[scale=0.5]{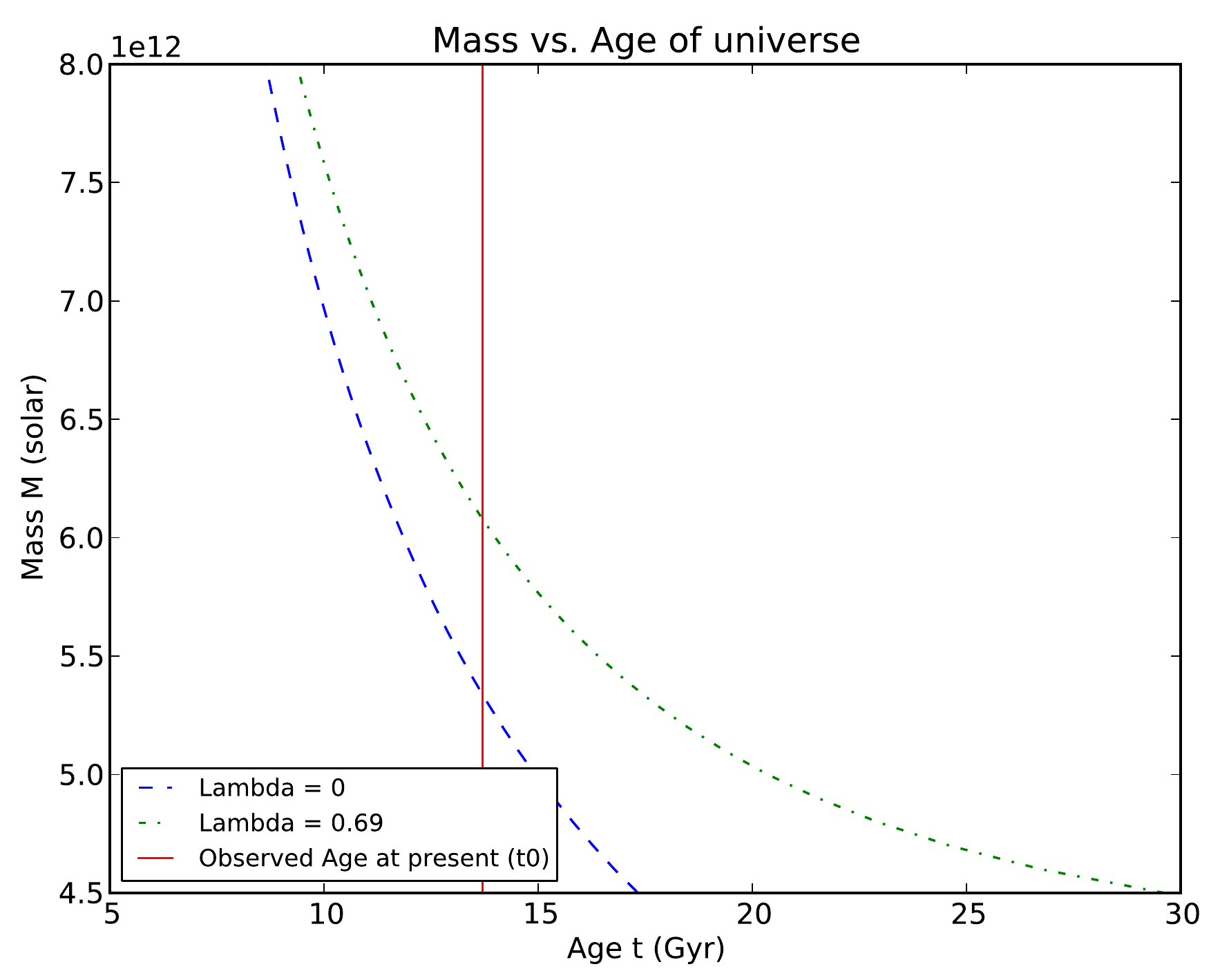}
\caption{For assumed $r, v, t$, and $\Omega_{\Lambda}$ as given in the text, the classical TA mass estimate is $M_{TA}=5.30 \pm 0.47 \times10^{12}\,\text{M}_{\sun}$, as compared with an effective TA mass that incorporates Dark Energy of $M_{TA \Lambda}=5.95 \pm 0.52 \times10^{12}\,\text{M}_{\sun}$ }
\label{fig:massvtime}  
\end{figure}

This is illustrated in Fig.\,\ref{fig:massvtime}, which is similar to Fig. 1 of \cite{chernin2009} and \cite{binneytremaine} (note that their input infall parameter and assumed cosmology are somewhat different from ours), who found an increased mass estimate of 15\%, which is similar to our findings. 
To derive the most up-to-date mass of the LG, we use now the latest infall values of $r  = 770 \pm 40$\,kpc  and $v=-109.3 \pm 4.4\,\textrm{km}\,\textrm{s}^{-1}$ from \cite{vdm12} to calculate the mass estimate with and without Dark Energy.  Using this $r$ and $v$, we get $\mta = (4.17\pm0.89) \times 10^{12}\,\text{M}_{\sun}$, and $\mtade = (4.73 \pm 1.03) \times 10^{12}\,\text{M}_{\sun}$, or a mass increase of 13\%.

The physical interpretation of this result is that due to the repulsive nature of Dark Energy, it would take more  mass to overcome the outward push from Dark Energy in order for the system to find itself in the configuration that we see today.

\section{Simulations}

\begin{figure} 
\centering 
\includegraphics[scale=0.4]{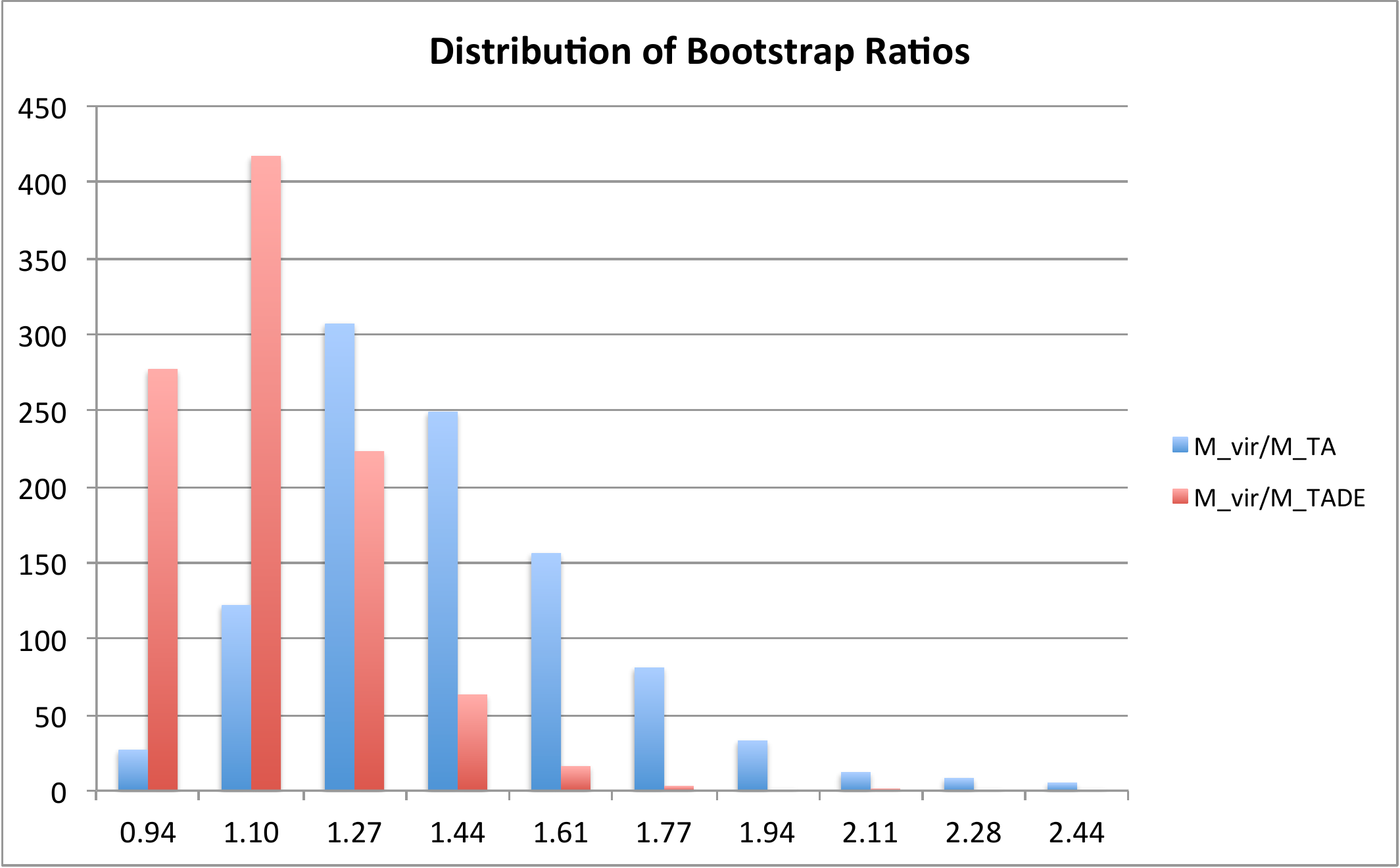}
\caption{Histogram showing the distribution of the average mass ratios found by bootstrapping the 16 simulation pairs for 1000 times.}
\label{fig:bootstraphist}  
\end{figure}


The TA model is extremely simplistic. While we have modified it above to incorporate $\Lambda$, there are other effects not taken into account in this model, such as tidal forces, non-radial motions etc. The only way to verify its validity  is by testing it against N-body simulations.We used galaxy pairs that were generated by the CLUES simulation which was run by \cite{CLUES}.  The galaxy pairs chosen were those that resembled the Local Group in terms of radius, virial mass, and approach velocity.  These galaxy pairs were generated from simulations using Gadget-2 code, using the cosmological parameters from WMAP5 (\cite{wmap5year}).  For our TA analysis of these simulation pairs, we assumed parameters close to those of WMAP5 for consistency, as opposed to the analysis we did for the LG, where we used current Planck 2013 values.

From the simulation data, 16 galaxy pairs were selected that displayed similar morphology to that of the LG: the pairs have similar masses, radial distances, and velocities when compared to M31 and the Milky Way. The data for each simulated galaxy pair was used to set the initial conditions for our radial differential equation model.  The TA masses were then evaluated for both the Dark Energy and the classical ($\Lambda = 0$) cases. The resulting mass estimates were then compared to the virial masses that were determined from the simulation data to see if the addition of the Dark Energy term had a significant effect on the TA mass estimations.

As can be seen in Table \ref{simtable}, the resulting TA mass estimates are consistently higher when the influence of $\Lambda$ is included in the TA model.  The mass estimates including Dark Energy led to a TA mass ratio of $\frac{\langle\mtade\rangle}{\langle\mta\rangle} = 1.13$, or an average increase of 13\% over classical mass estimates. This is in accord with the 13\% mass increase that we found for the LG when using input parameters from \cite{vdm12}.

\begin{figure} 
\centering 
\includegraphics[scale=0.4]{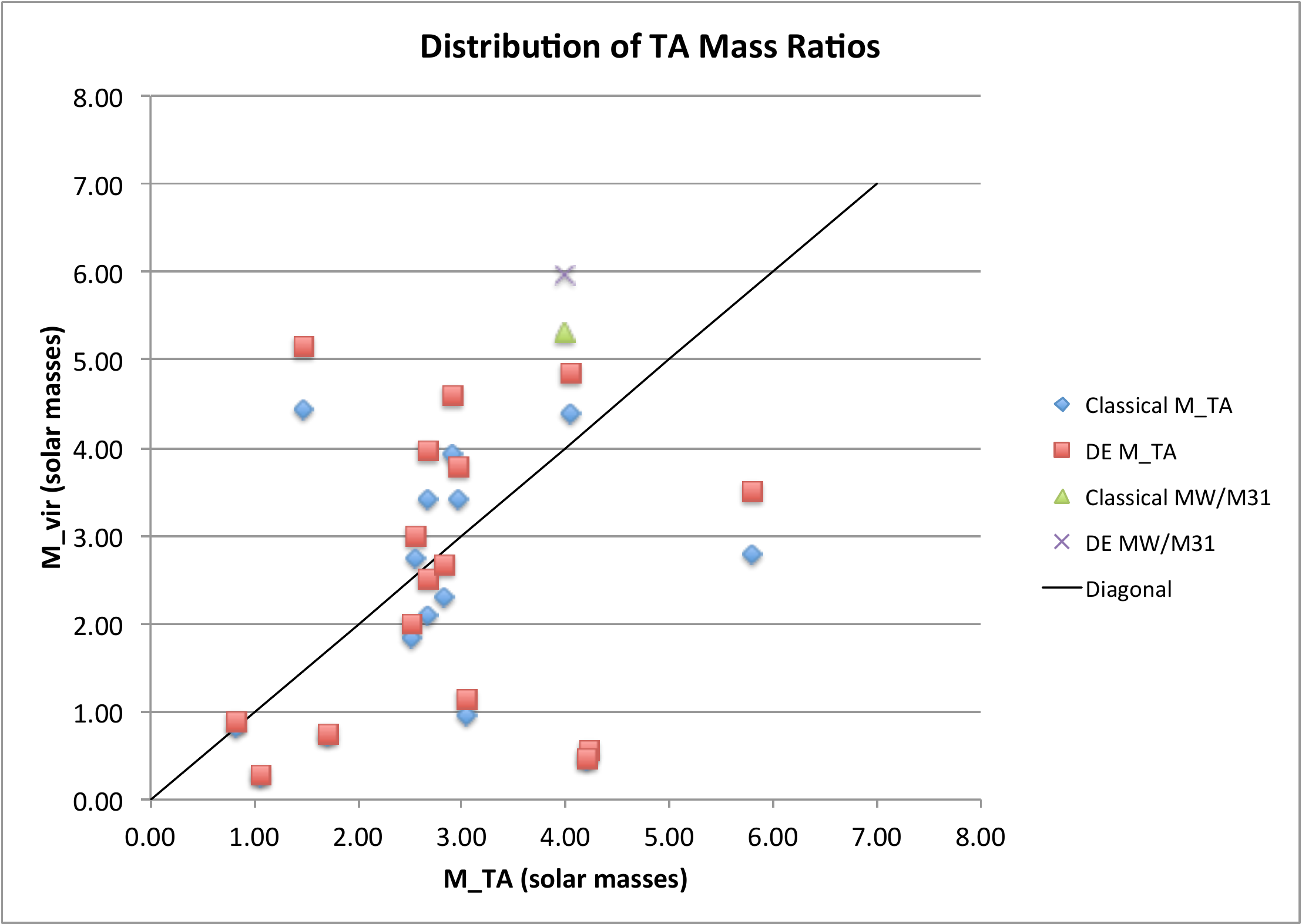}
\caption{The distribution of TA mass estimates, showing that  as mass increases, the difference between the classical $M_{\textrm{TA}}$ and $M_{\textrm{TA}\Lambda}$ also increases.}
\label{fig:massratios}  
\end{figure}

\begin{table}
  \centering
	\begin{tabular}{| l | c | c | c | c | c | c |}
	\hline
	Pair ID & $M_{vir}$  & $M_{\textrm{TA}}$ & $M_{\textrm{TA}\Lambda}$  \\ \hline \hline
	\textbf{LG} & \textbf{4.0} & \textbf{5.30} & \textbf{5.95}  \\ 
	1 & 5.80 & 2.78 & 3.50  \\ 
 	2 & 4.24 & 0.51 & 0.55  \\ 
	3 & 4.22 & 0.43 & 0.47 \\ 
 	4 & 4.05 & 4.39 & 4.84  \\ 
 	5 & 3.04 & 0.96 & 1.14  \\ 
 	6 & 2.91 & 3.91 & 4.58  \\ 
 	7 & 2.67 & 3.42 & 3.97  \\ 
 	8 & 2.53 & 1.84 & 1.99  \\ 
 	9 & 2.55 & 2.74 & 2.98 \\ 
	10 & 2.97 & 3.40 & 3.79  \\ 
	11 & 1.72 & 0.73 & 0.75 \\ 
	12 & 2.69 & 2.09 & 2.51  \\ 
	13 & 2.84 & 2.30 & 2.66  \\ 
	14 & 1.47 & 4.43 & 5.15 \\ 
	15 & 1.07 & 0.26 & 0.28  \\ 
	16 & 0.83 & 0.81 & 0.89  \\ 
	\hline
	\end{tabular}
	\caption{The virial masses and calculated TA masses, both with and without Dark Energy, for the LG and the 16 LG-like pairs from the CLUES simulation.  Masses are given in units of $10^{12}\,\text{M}_{\sun}$. }
\label{simtable}
\end{table}

We wish now to use the simulations to calibrate the derived TA masses with the virial masses, and to quantify the systematic uncertainties.  In order to check the robustness to outliers (e.g. pair number 14) , the 16 pairs were used in a bootstrap analysis, where 1000 sets of 16 pairs were created allowing repetition, and the average mass ratios found for each of the 1000 sets.  The distribution of the average values for the ratios of virial mass to classical TA mass  $\frac{\langle\mvir\rangle}{\langle\mta\rangle}$ and virial mass to Dark Energy TA mass $\frac{\langle\mvir\rangle}{\langle\mtade\rangle}$ is shown in the histogram in Fig. \ref{fig:bootstraphist}. The overall average mass ratios from the bootstrap analysis were $\frac{\langle\mvir\rangle}{\langle\mta\rangle} = 1.34 \pm 0.26$ and $\frac{\langle\mvir\rangle}{\langle\mtade\rangle} = 1.04 \pm 0.16$. The average ratio of Dark Energy TA mass to classical TA mass from the bootstrap analysis was $\frac{\langle\mtade\rangle}{\langle\mta\rangle} = 1.29 \pm 0.15$ (and similarly for the median). This somewhat differs from the above ratio of 13\% which is a reflection of the large scatter in the data.

In \cite{timingargwhite}, they analysed LCDM N-body simulations and found a calibration ratio of TA mass to $M_{200}$ mass (which somewhat differs from $\mvir$) for the LG to be $\frac{M_{200}}{\mta} \approx 1.6$
where $M_{200} = 5.27 \times10^{12}\,\text{M}_{\sun}$.  In their paper, they only considered the classical TA and pointed out this discrepancy between the masses. Similarly, Bar Asher \& Hoffman (unpublished 2010) found a ratio of $\mvir / \mta = 1.55 \pm  0.26$ from their LCDM CLUES simulations.  We propose that most of the shift can be explained by the fact the classical TA model has historically neglected to include Dark Energy.

\section{Discussion}

%
In this study we have modified the classical TA to include the effect of the Cosmological Constant.  We find in the case of the LG (using values from \cite{vdm12}) and the latest parameters from Planck that the TA mass is ~13\% higher when $\Lambda$ is included in the TA. This agrees with earlier contributions from \cite{chernin2009}. Lynden-Bell (2013) derived an approximate analytical solution that gives a very similar mass increase. We believe that this explains in part the discrepancy found by \cite{timingargwhite} between their LCDM simulations and the classical TA (in the absence of Dark Energy). We conclude that this TA example illustrates that in physical coordinates $\Lambda$ has imprints on Megaparsec scales.

When we repeat our TA analysis with and without Dark Energy over 16 LG-like pairs from the CLUES simulation, we find an average increase of 13\% over classical TA masses. Applying it to the LG mass estimation we find a calibrated $\mvir = (4.92 \pm 1.08 \textrm{(obs.)} \pm 0.79 \textrm{(sys.)}  )\times 10^{12} M_{\sun}$. However, even with the inclusion of $\Lambda$, the TA may not capture the reality of nature because of tidal effects (\cite{TAtidal}), and also because the objects are not point-like. Additionally, in the TA, all motion is assumed to be purely radial, ignoring any transverse motion. Despite the complexity of the problem, we find that the systematic effects account to only 16\%, compared with the 21\% attributed to the observational errors on the input parameters.  This conclusion should be further investigated with a larger number of simulation pairs, and with a wider range of Dark Energy models.  

The Timing Argument by \cite{kahnwoltjer1959} was among the first methods that illustrated the dark matter problem (e.g. \cite{freeman2013} for historical perspective).  Given the latest results on baryon fraction from cosmology, where $\Omega_b/\Omega_m = 0.185 \pm 0.003$, we can speculate that under the simplistic assumption that the LG baryon fraction is the same as the universal one, then $(0.910 \pm 0.015) \times 10^{12}\,\text{M}_{\sun}$ of the LG virial mass is in baryonic form, where the erros bar here is just due to the uncertainty in the universal baryonic fraction.

\section{Acknowledgements}
We thank O. Host for his help with the numerical  work, D. Lynden-Bell for deriving an analytic approximation to check our numerical results, and the CLUES Team and S. Bar-Asher for their contribution to the simulation analyses. We also acknowledge helpful discussion with J. Frieman, M. Milgrom and S. White. OL acknowledges a Royal Society Wolfson Research Merit Award, a Leverhulme Senior Research Fellowship and an Advanced Grant from the European Research Council. YH has been supported by the ISF (1013/12).

\bibliographystyle{mn2e}
\bibliography{timingargbib}

\label{lastpage}

\end{document}